\documentclass[12pt,cite,epsf,epsfig]{article}
\usepackage{epsfig}
\title{CP violation in the decay
mode $B\rightarrow \pi \gamma \gamma $}
\author{S.~R.~Choudhury$^a$\thanks{E--mail : src@physics.du.ac.in},
G.~C.~Joshi$^b$\thanks{E--mail: joshi@tauon.ph.unimelb.edu.au},
Namit Mahajan$^c$\thanks{E--mail : nmahajan@mri.ernet.in} and
B.~H~.~J.~McKellar$^b$\thanks{E--mail:b.mckellar@physics.unimelb.edu.au}\\
	$a.$~{\em Department of Physics and Astrophysics,} \\
	 {\em University of Delhi, Delhi-110 007, India.}\\
	$b.$~{\em School of Physics, University of Melbourne,}\\
	{\em   Australia.}\\
        $c.$~{\em Harish-Chandra Reserach Institute,}\\
          {\em Chhatnag Road, Jhunsi, Allahabad - 211019, India.}}

\def\be{\begin{equation}}
\def\ee{\end{equation}}
\def\bea{\begin{eqnarray}}
\def\eea{\end{eqnarray}}

\begin{document}
\maketitle

\begin{abstract}
Within the framework of Standard Model, the exclusive decay mode
$B\rightarrow \pi \gamma \gamma $ is studied.
  Although the usual short distance
contribution is small compared to the similar $B\rightarrow K\gamma\gamma$
mode, the process offers the possibility of studying the CP violation,
a feature absent in the $B \rightarrow K$ counterpart. \\
{\bf Keywords}:Rare B decay \\
{\bf PACS}:13.25.Hw, 13.40.Hq
\end{abstract}

\begin{section}{Introduction}
As has been emphasised time and again, because of very clean experimental
signatures,
radiative decays of the B-meson can serve to be very useful tools to
test the underlying structure of the theory in question responsible
for flavour changing neutral currents (FCNC). The radiative decays,
in particular, are very sensitive to higher order QCD corrections
and any new physics effects beyond the Standard Model (SM).
The inclusive decay mode $B\rightarrow X_s\gamma$ as well as
the exclusive mode $B\rightarrow K^*\gamma$ have been experimentally
measured and have invited a lot of theoretical attention.
It is expected that the future B-factories should be able to
make measurements of two photon modes. The quark level decay,
$b\rightarrow s\gamma\gamma$ has been studied in \cite{lin}-\cite{reina}.
The basic amplitude in question receives contributions from the irreducible
triangle diagram and the reducible pieces (where the second photon
is emitted by one of the external quark lines of a basic $b\to s\gamma$
amplitude). The quark level
amplitude is appropriate for the inclusive two photon process. For
an exclusive channel, with some meson M in the final state,
the reducible diagram is to be thought of as the one in which the second
photon is attached to one of the external meson legs of the basic
amplitude $ B \rightarrow M \gamma$ rather than the quark level process.
It is then straighforward to note that this amplitude vanishes if the meson M
in question is scalar or pseudoscalar. Therefore, for such cases, it suffices
to consider the irreducible contribution only. Apart from these
short distance contributions, one can have the long distances contributions
as well. Of course, in the case of resonances contributing to any
process, the resonant contribution is a reducible one. However, in the
above, the demarcation between reducible and irreducible diagrams is
more specific in that it is used for the usual short distance,
non-resonant contribution only.
The two photon process proceeding via $b \rightarrow s$ transition
has been recently studied \cite{chou1} and it has been pointed out that
the long distance contributions can be of the same order, in fact sometimes
  larger, compared to the short distance irreducible contributions.\\

The present study extends  these ideas to the decay
$B\rightarrow \pi \gamma \gamma $
proceeding via the transition $b \rightarrow d$. The amplitude for this
mode suffers from CKM suppression when compared with the
$B\rightarrow K \gamma \gamma $ amplitude. However, in contrast to
the $B\rightarrow K$ process, $B\rightarrow \pi$ channel offers the
possibility of studying the CP asymmetry.
We estimate the branching ratio and CP asymmetry for the
process $ B \rightarrow \pi \gamma \gamma $ taking account of the irreducible
contribution as well as all significant resonance contributions.

\end{section}

\begin{section}{Effective Hamiltonian and Irreducible triangle contribution}
The effective Hamitonian relevant for a $b\to d$  transition contributes 
to radiative processess through  triangle diagrams with the photon being 
emitted by the
  quark loop and has the form \cite{sehgal}
\bea
{\cal{H}}_{eff} &=& -\frac{G_F}{\sqrt{2}}V_{td}^*V_{tb}\Bigg[\sum_i
                          C_i(\mu)O_i(\mu)\\ \nonumber
&-& \lambda_u[C_1(O_1^u(\mu) - O_1(\mu)) + C_2(O_2^u(\mu) - O_2(\mu))]\Bigg]
\eea
with
\[
O_1 = (\bar{d}_ic_j)_{V-A}(\bar{c}_jb_i)_{V-A},
\]
\[
O_2 = (\bar{d}_ic_i)_{V-A}(\bar{c}_jb_j)_{V-A},
\]
\[
O_3 = (\bar{d}_ib_i)_{V-A}\sum_q(\bar{q}_jq_j)_{V-A},
\]
\[
O_4 = (\bar{d}_ib_j)_{V-A}\sum_q(\bar{q}_jq_i)_{V-A},
\]
\be
O_5 = (\bar{d}_ib_i)_{V-A}\sum_q(\bar{q}_jq_j)_{V+A},
\ee
\[
O_6 = (\bar{d}_ib_j)_{V-A}\sum_q(\bar{q}_jq_i)_{V+A},
\]
\[
O_7 = \frac{e}{16\pi^2}\bar{d}_i\sigma^{\mu\nu}(m_dP_L +
m_bP_R)b_iF_{\mu\nu},
\]
\[
O_8 = \frac{g}{16\pi^2}\bar{d}_i\sigma^{\mu\nu}(m_dP_L + m_bP_R)
            T_{ij}^ab_jG_{\mu\nu}^a.
\]
and
\be
O_1^u = (\bar{d}_iu_j)_{V-A}(\bar{u}_jb_i)_{V-A}
\ee
\be
O_2^u = (\bar{d}_iu_i)_{V-A}(\bar{u}_jb_j)_{V-A},
\ee
In writing the effective Hamiltonian, the unitarity of the CKM matrix
has been used and the parameter $\lambda_u$ is:
\be
\lambda_u = \frac{V_{ud}^*V_{ub}}{V_{td}^*V_{tb}}
\ee
In the Wolfenstein parametrization of the CKM matrix, the above equation simply
reads
\be
\lambda_u = \frac{\rho(1-\rho) - \eta^2}{(1-\rho)^2 + \eta^2}
-\imath\frac{\eta}{(1-\rho)^2 + \eta^2} + ......
\ee

A simple Fierz rearrangement of $O_1^u$ and $O_2^u$ yields,
\be
O_1^u = -(\bar{d}_ib_i)_{V-A}(\bar{u}_ju_j)_{V-A} = -O_3(q=u)
\ee
and
\be
O_2^u = -(\bar{d}_ib_j)_{V-A}(\bar{u}_ju_i)_{V-A} = -O_4(q=u)
\ee

The effecive Hamiltonian therefore becomes
\bea
{\cal{H}}_{eff} &=& -\frac{G_F}{\sqrt{2}}V_{td}^*V_{tb}\Bigg[
C_1O_1(\mu)[1+\lambda_u] + C_2O_2(\mu)[1+\lambda_u]\\ \nonumber
&+& O_3[C_3+C_1\lambda_u \delta_{uq}] + O_3[C_4+C_2\lambda_u \delta_{uq}]\\
\nonumber
&+& \sum_{i>4} C_i(\mu)O_i(\mu)\Bigg]
\eea
In the above equation, $\delta_{uq}$ simply implies that this term contributes
only for the u-quark and there is no summation over repeated indices.\\

The quark level transition amplitude for
$b\longrightarrow d \gamma\gamma$ (with an incoming $b$ line and an
outgoing
$d$ line and the two photons being emitted by the quark loop) is
\bea
{\cal{M}}_{b\rightarrow d} &=& \Bigg[\frac{16\sqrt{2}\alpha
G_FV_{td}^*V_{tb}}{9\pi}
            \Bigg]\bar{u}(p_d)\Bigg\{\sum_q
A_qJ(m_q^2)\gamma^{\rho}P_L
R_{\mu\nu\rho}\\ \nonumber
&+& \iota B(m_dK(m_d^2)P_L + m_bK(m_b^2)P_R) T_{\mu\nu}\\ \nonumber
&+& C(-m_dL(m_d^2)P_L + m_bL(m_b^2)P_R)\epsilon_{\mu\nu\alpha\beta}
k_1^{\alpha}k_2^{\beta}\Bigg\}u(p_b)\epsilon^{\mu}(k_1)\epsilon^{\nu}(k_2),
\eea
where
\[
R_{\mu\nu\rho} =
{k_1}_{\nu}\epsilon_{\mu\rho\sigma\lambda}k_1^{\sigma}
k_2^{\lambda} -
{k_2}_{\mu}\epsilon_{\nu\rho\sigma\lambda}k_1^{\sigma}
k_2^{\lambda} + (k_1.k_2)\epsilon_{\mu\nu\rho\sigma}(k_2-k_1)^{\sigma},
\]
\[
T_{\mu\nu} = {k_2}_{\mu}{k_1}_{\nu} - (k_1.k_2) g_{\mu\nu},
\]
\[
A_u = 3(C_3+C_1\lambda_u-C_5) + (C_4+C_2\lambda_u-C_6),
\]
\be
A_d = \frac{1}{4}[3(C_3-C_5) + (C_4-C_6)],
\ee
\[
A_c = 3(C_1(1+\lambda_u)+C_3-C_5) + (C_2(1+\lambda_u)+C_4-C_6),
\]
\[
A_s = A_b = \frac{1}{4}[3(C_1(1+\lambda_u)+C_3-C_5) +
(C_2(1+\lambda_u)+C_4-C_6)],
\]
and
\[
B = C = -\frac{1}{4}(3C_6+C_5).
\]

The quark level amplitude when sandwiched between the hadronic states
gives the appropriate matrix element describing the meson decay. We thus get,
\bea
{\cal{M}}_{\rm{irred}}(B\rightarrow \pi\gamma\gamma) &=&
\Bigg(\frac{16\sqrt{2}\alpha G_FV_{td}^*V_{tb}}{9\pi}
            \Bigg)\Bigg[\frac{1}{2}\langle
\pi\vert\bar{d}\gamma^{\rho}b
\vert B\rangle\sum_q A_qJ(m_q^2)R_{\mu\nu\rho} \nonumber \\
&+&\frac{1}{2}\langle \pi\vert \bar{d}b\vert B\rangle\Bigg\{
\iota B(m_dK(m_d^2) + m_bK(m_b^2)) T_{\mu\nu}\\ \nonumber
&+& C(-m_dL(m_d^2) + m_bL(m_b^2))\epsilon_{\mu\nu\alpha\beta}
k_1^{\alpha}k_2^{\beta}\Bigg\}\Bigg]\epsilon^{\mu}(k_1)\epsilon^{\nu}(k_2)
\eea
In the above expressions we introduced the functions
\[
J(m^2) = I_{11}(m^2), \hskip 1cm K(m^2) = 4I_{11}(m^2) - I_{00}(m^2),
\hskip 1cm     L(m^2) = I_{00}(m^2),
\] where
\be
I_{pq}(m^2) = \int_0^1dx\int_0^{1-x}dy \frac{x^py^q}{m^2 -
2(k_1.k_2)xy - \iota
\epsilon}
\ee
  We use the following parametrization for the matrix elements of the
  quark vector current:
\be
\langle \pi^-(p)\vert(\bar{d}u)_{V-A}\vert 0\rangle = \imath f_{\pi}p_{\mu}
\ee
\bea
\langle \pi^-\vert\bar{d}\gamma_{\mu}b \vert B^-\rangle &=&
\Bigg((p_B+p_{\pi})_{\mu}
- \frac{m_B^2-m_{\pi}^2}{q^2}q_{\mu}\Bigg)F_1^{B\pi}(q^2) \\ \nonumber
&+& \Bigg(\frac{m_B^2-m_{\pi}^2}{q^2}\Bigg)q_{\mu}F_0^{B\pi}(q^2),
\eea
with $q=p_b-p_{\pi}=k_1+k_2$. It then follows that
\bea
\langle \pi^-\vert\bar{d}b \vert B^-\rangle &=& (m_b-m_d)^{-1}
\Bigg[q_{\mu}\langle \pi\vert\bar{d}\gamma^{\mu}b \vert
  B\rangle \Bigg]\\ \nonumber
&=& (m_b-m_d)^{-1} (m_B^2-m_{\pi}^2)F_0^{B\pi}(q^2).
\eea
Other matrix elements can be related to these using the isospin relations:
\[
\langle \pi^-(p)\vert(\bar{d}u)_{V-A}\vert 0\rangle =
\sqrt{2}\langle \pi^0(p)\vert(\bar{u}u)_{V-A}\vert 0\rangle =
-\sqrt{2}\langle \pi^-(p)\vert(\bar{d}d)_{V-A}\vert 0\rangle
\]
and
\[
\langle \pi^-\vert\bar{d}\gamma_{\mu}b \vert B^-\rangle =
\langle \pi^+\vert\bar{u}\gamma_{\mu}b \vert \bar{B^0}\rangle =
\sqrt{2}\langle \pi^0\vert\bar{d}\gamma_{\mu}b \vert \bar{B^0}\rangle =
\sqrt{2}\langle \pi^0\vert\bar{u}\gamma_{\mu}b \vert B^-\rangle
\]

We use the explicit
numerical dependence of $F(q^2)$ on $q^2$ given by Cheng \emph{et al}
\cite{cheng}.
\end{section}
\begin{section}{Resonance contributions}
\subsection{The $\eta_c$, $\eta$ and $\eta^{\prime}$  resonances}
  The $\eta_c$
contribution comes via the decay $B\rightarrow
\pi \eta_c$,
  followed by $\eta_c$ decaying into two photons.

The amplitude for $\eta_c$ to decay to two photons is
parametrized
as \cite{reina}
\be
\langle\gamma\gamma\vert T\vert\eta_c\rangle = 2\imath B_{\eta_c}
\epsilon^{\mu\nu\alpha\beta}{\epsilon_1}_{\mu}^*{\epsilon_2}_{\nu}
^*
{k_1}_{\alpha}{k_2}_{\beta}. \label{eq1}
\ee
The coefficient $B_{\eta_{c}}$ can be determined from the
$\eta_{c} \to \gamma \gamma $ decay rate.
\[
\Gamma (\eta_{c}\rightarrow\gamma\gamma) = \frac{\vert B_{\eta}\vert^2
m_{\eta_c}^3}
{16\pi}.
\]
The $B \to \pi \eta_{c}$ amplitude
\[
\langle \pi\eta_c\vert T\vert B\rangle = -\langle \pi\eta_c\vert
{\cal{H}}_{\rm{eff}}
\vert B\rangle \, ,
\]
can be got from the relevant piece of
  ${\cal{H}}_{\rm{eff}}$, which after Fierz transformation reads
\be
{\cal{H}}_{\rm{eff}} =
-\frac{G_F}{\sqrt{2}}V_{cb}V_{cd}^*(C_1+\frac{C_2}{3})
(\bar{c}c)_{V-A}(\bar{d}b)_{V-A}.
\ee
Using factorization and the following definitions
\[
\langle 0\vert A_{\mu}^c\vert\eta_c\rangle \equiv \imath
f_{\eta_{c}}q_{\mu}
\hskip 1cm A_{\mu}^c = \bar{c}\gamma_{\mu}\gamma_5 c
\]
\[
q^{\mu}\langle \pi^-\vert\bar{d}\gamma_{\mu}b\vert B\rangle =
F_0(m_{\eta_c}^2)
(m_B^2-m_{\pi}^2)
\]
we can write
\be
\langle \pi^-\eta_c\vert T\vert B\rangle =
-\imath\frac{G_F}{\sqrt{2}}V_{cb}V_{cd}^*(C_1+\frac{C_2}{3})
\Bigg(f_{\eta_c}F_0(m_{\eta_c}^2)(m_B^2-m_{\pi}^2) -
f_{\pi}F_0(m_{\pi}^2)(m_B^2-m_{\eta_c}^2)\Bigg)
\ee
where a dipole form of the form factor $F_0(m^2)$ is used.

  The $\eta_c$ resonance contribution is thus,
\bea {\cal{M}}_{\eta_c} &=&
2B_{\eta_c}\frac{G_F}{\sqrt{2}}
V_{cb}V_{cd}^*(C_1+\frac{C_2}{3})\\ \nonumber
&&\Bigg[f_{\eta_c}
F_0(m_{\eta_c}^2)(m_B^2-m_{\pi}^2) -
f_{\pi}F_0(m_{\pi}^2)(m_B^2-m_{\eta_c}^2)\Bigg]
\label{eq-eta-c}\\
\nonumber && {\epsilon_1^*}^{\mu}(k_1){\epsilon_2^*}^{\nu}(k_2)
\epsilon_{\mu\nu\alpha\beta}k_1^{\alpha}k_2^{\beta}
\frac{1}{q^2-m_{\eta_c}^2+ \imath
m_{\eta_c}\Gamma_{\rm{total}}^{\eta_c}}. \eea

Analogous to  $\eta_c$, $\eta^{\prime}$ can also contribute
via the effective Hamiltonian, eq.(15). However, unlike $\eta_c$,
$\eta^{\prime}$ has a very strong coupling to a two gluon state.
A number of theoretical models for $\eta^{\prime}$ production in B-decays
via two gluon states have been proposed \cite{ali,ahmady1,ahmady2}. But these
theoretical models have their own uncertainities attached to them. Instead of
relying on any such model, we directly use the experimental data
for $B\rightarrow \pi\eta^{\prime}$ decay process \cite{exp1} and parametrise
the $\eta^{\prime}$ resonance contribution as ($i=-,0$)
\be
{\mathcal{M}}_{\eta^{\prime}} = 2 B_{\eta^{\prime}}F(B^i\pi^i\eta^{\prime})
{\epsilon_1^*}^{\mu}(k_1){\epsilon_2^*}^{\nu}(k_2)
\epsilon_{\mu\nu\alpha\beta}k_1^{\alpha}k_2^{\beta}
\frac{1}{q^2-m_{\eta^{\prime}}^2+ \imath
m_{\eta^{\prime}}\Gamma_{\rm{total}}^{\eta^{\prime}}} \label{eq.etap}
\ee
where $B_{\eta^{\prime}}$ is defined as for $\eta_c$ with
$\eta_c\rightarrow\eta^{\prime}$ and the factor
  $F(B^i\pi^i\eta^{\prime})$ is related
to the decay rate $\Gamma(B^i\rightarrow \pi^i\eta^{\prime})$ as:
\be
\Gamma(B^i\rightarrow \pi^i\eta^{\prime}) = \frac{1}{16\pi m_B}
\vert F(B^i\pi^i\eta^{\prime})\vert^2
\lambda^{\frac{1}{2}}(1,\frac{m_{\pi}^2}{m_B^2},\frac{m_{\eta^{\prime}}^2}
{m_B^2})
\ee
\indent The branching ratio for $B\to \pi \eta$ has also been measured
\cite{exp2} and is found to be larger than the $\eta^{\prime}$
branching ratio. Moreover, the branching ratio for $\eta$ to go into
two photons is roughly $20$ times than that of
$\eta^{\prime}$. Therefore, $\eta$ channel is expected to contribute
the largest. Again, the amplitude can be directly read off from
Eq.(\ref{eq.etap}) by replacing the $\eta^{\prime}$ quantities by
appropriate $\eta$ values.
\subsection{Contribution due to $\rho$}
The $\rho$ meson contributes to the process in the following way:

\[
B^i(p_B) \longrightarrow {\rho}^i + \gamma (k_1), \hskip 1.5cm i~=~\pm,~0
\]
followed by
\[
{\rho}^i \longrightarrow \pi^i + \gamma(k_2),
\]
and the process with $k_1 \leftrightarrow k_2$.

It has been emphasized \cite{stech} that the weak-annihilation contribution
to $B \to \rho \gamma$ can be significant and thus we take into account this
effect also while writing the $B \to \rho \gamma$ vertex. Therefore,
\bea
\langle\rho(p_{\rho})\gamma(k_1)\vert T\vert B(p_B)\rangle &=&
-\frac{eG_F}{\sqrt{2}}V_{tb}V_{td}^*\epsilon^{*\nu}(k_1)
\epsilon^{*\beta}(p_{\rho}))\\ \nonumber
&&\Bigg[F_{PC}^{Total}\epsilon_{\mu\nu\alpha\beta}k_1^{\mu}p_{\rho}^{\alpha} +
\imath F_{PV}^{Total}(g_{\nu\beta}p_B.k_1 - {p_B}_{\nu}{k_1}_{\beta})\Bigg]
\eea

where, $F_{PC}^{Total}$ and $F_{PV}^{Total}$ are the parity conserving and
parity violating form factors including the weak annihilation contributions as
given in \cite{stech}. We parametrize the $\rho^i\to \pi^i\gamma$ transition as
\be
\langle\pi(p_{\pi})\gamma(k_2)\vert T\vert \rho(p_{\rho})\rangle =
g^i_{\rho\pi\gamma}\epsilon^{\delta}(p_{\rho})
\epsilon^{*\sigma}(k_2)k_2^{\kappa}
p_{\rho}^{\lambda}\epsilon_{\kappa\sigma\lambda\delta}
\ee
and $g^i_{\rho\pi\gamma}$ is determined from the corresponding decay rate.

Therefore, the complete $T$ matrix element for the $\rho$ resonance can be
written as
\bea
\langle \pi\gamma\gamma\vert T \vert B\rangle &=& {\cal{M}}_{\rho} \\ \nonumber
&=& -\frac{eG_F}{\sqrt{2}}V_{tb}V_{td}^*g^i_{\rho\pi\gamma}
{\epsilon^*}_{\mu}(k_1){\epsilon^*}_{\nu}(k_2)\\ \nonumber
&&\Bigg\{\epsilon^{\alpha\nu\gamma\delta}
{k_2}_{\alpha}(p_B-k_1)_{\gamma}{k_1}_{\beta^{\prime}}
\Bigg[\frac{\Bigg(g_{\delta\sigma^{\prime}}-\frac{(p_B-k_1)_
{\delta}(p_B-k_1)_{\sigma^{\prime}}}{m_{K^*}^2}\Bigg)}
{(p_B-k_1)^2-m_{K^*}^2+\imath m_{K^*}\Gamma^{K^*}_{\rm{total}}}
\Bigg]\\\nonumber
&&\Bigg[\imath F_{PC}^{Total}\epsilon^{\mu\beta^{\prime}\sigma^{\prime}
\tau^{\prime}}
(p_B-k_1)_{\tau^{\prime}} - F_{PV}^{Total}
(g^{\mu\sigma^{\prime}}(p_B-k_1)^{\beta^{\prime}}
-g^{\beta^{\prime}\sigma^{\prime}}(p_B-k_1)^{\mu})\Bigg] \\ \nonumber
&& + ~\Bigg(\begin{array}{cl}
\mu \leftrightarrow \nu \\
k_1 \leftrightarrow k_2
\end{array}\Bigg)
\Bigg\}
\eea
\subsection{$\omega$, $\phi$ and $J/\psi$ contributions}
The mesons $\omega$, $\phi$ and $J/\psi$ contribute only to the neutral 
decay mode.
The matrix elements for the $\omega$ and $\phi$ can be easily got from that
of $\rho$ by making appropriate changes to quantities corresponding to
$\omega$ and $\phi$ and thus we don't write the explicit expressions here.\\
The neutral mode also receives some contribution from $J/\psi$ via the
annihilation diagram, which we expect to be small and do not include.

\end{section}

\begin{section}{Results}
It is worth mentioning that apart from the $\eta_c$ contribution, there is no
handle for determing the relative sign of other matix elements. This introduces
some amount of uncertainity in the result for the total decay rate.
The differential decay rate is given by
\be
\frac{d\Gamma}{d\sqrt{s_{\gamma\gamma}}} = \Bigg(\frac{1}{512m_B\pi^3}\Bigg)
\sqrt{s_{\gamma\gamma}}
\Bigg[\Bigg(1 - \frac{s_{\gamma\gamma}}{m_B^2} + \frac{m_{\pi}^2}
{m_B^2}\Bigg)^2
  - \frac{4m_{\pi}^2}{m_B^2}\Bigg]^{\frac{1}{2}}\int_0^{\pi}d\theta~\sin\theta
  \vert{\mathcal{M}}_{tot}\vert^2
\ee
where ${\mathcal{M}}_{tot}$ represents the complete matrix elemnt
for the decay process including the resonance terms,
$\sqrt{s_{\gamma\gamma}}$ is the C.M.
energy of the two photons
and $\theta$ denotes the angle which the decaying B-meson makes
with one of the
two photons in the $\gamma\gamma$ C.M. frame.
Figure 1 shows the differential decay rate as a function of the invariant
mass of the two photons for the charged mode. The $\eta$, $\eta^{\prime}$ and
$\eta_c$ peaks show up at the corresponding mass values while the
$\rho$ contribution gets spread over the whole range of
$\sqrt{s_{\gamma\gamma}}$.
Further, the interference effects between the resonant and the
irreducible contributions are very small, with the result that
the results are practically independent of the relative sign parameters that
one may have introduced for each of the resonance terms.

The total decay rate, including all resonance contributions 
  for this process is calculated to be :
\begin{equation}
Br( B^- \rightarrow \pi^- \gamma \gamma) \sim 1.7 \times 10^{-6}
\end{equation}
  We quote the individual contributions to the branching ratio:
\bea
\eta_c \mapsto   2 \times 10^{-9} & \quad\quad &
\rho \mapsto 2 \times  10^{-9} \\ \nonumber
\rm{Irreducible} \mapsto 3 \times 10^{-8} & \quad\quad & \rm{interference}
\mapsto 7 \times 10^{-9} \\ \nonumber
\eta \mapsto   1.5 \times 10^{-6} & \quad\quad &
\eta^{\prime} \mapsto 6 \times  10^{-8}
\eea
\vskip 2cm
\begin{figure}[ht]
\vspace*{-1cm}
\centerline{
\epsfxsize=10cm\epsfysize=10.0cm
                      \epsfbox{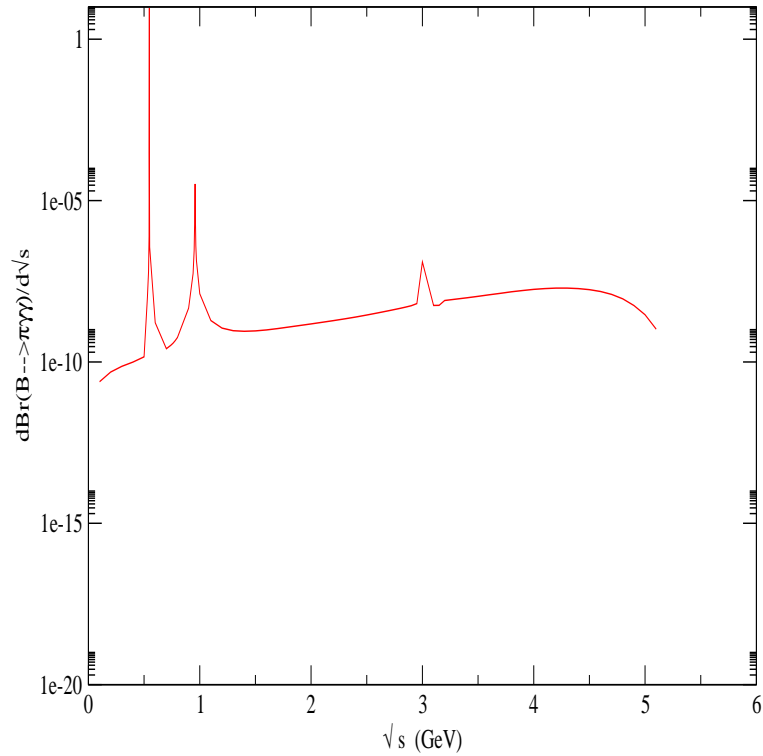}}
\caption{\em Our result for  the spectrum of $B^-\rightarrow \pi^-
\gamma\gamma$ plotted with logscale on y-axis.
  The parameters used are
listed in the appendix.
	}
\label{fig:fig1}
\end{figure}
The branching ratio for the neutral mode is also expected to be of
similar magnitude. The neutral mode gets additional,
although small, (${\mathcal{O}} \sim \rho$)-contributions from the
$\omega$ and $\phi$ modes as well. More accurate statements can be
made about the neutral mode once $B\to \pi\eta (\eta^{\prime})$ is
observed.   \\

As mentioned earlier, the process, although CKM suppressed, opens up the
possibility of studying the CP asymmetry. We define the CP asymmetry as
\be
{\mathcal{A}}_{CP} = \frac{\frac{d\Gamma}{d\sqrt{s_{\gamma\gamma}}} -
\frac{d\bar{\Gamma}}{d\sqrt{s_{\gamma\gamma}}}}
{\frac{d\Gamma}{d\sqrt{s_{\gamma\gamma}}} +
\frac{d\bar{\Gamma}}{d\sqrt{s_{\gamma\gamma}}}}
\ee
with $\Gamma$ and $\bar{\Gamma}$ denoting the decay rates for
$B^-$ and $B^+$ (or $B^0$ and $\bar{B^0}$) respectively.
The resonances by themselves do not contribute to CP asymmetry. The
CKM factors for them are dominantly real and in any case will be
overall multiplicative factors unchanged while going from $B$ to
$\bar{B}$ when calculating the individual resonance contributions.
However, there can be small contribution coming from the
interference between the resonant and irreducible amplitudes. However,
the interference terms being rather small have been neglected in this
analysis, which in principle can contribute to the asymmetry.
\vskip 2cm
\begin{figure}[ht]
\vspace*{-0.5cm}
\centerline{
\epsfxsize=10.0cm\epsfysize=10.0cm
                      \epsfbox{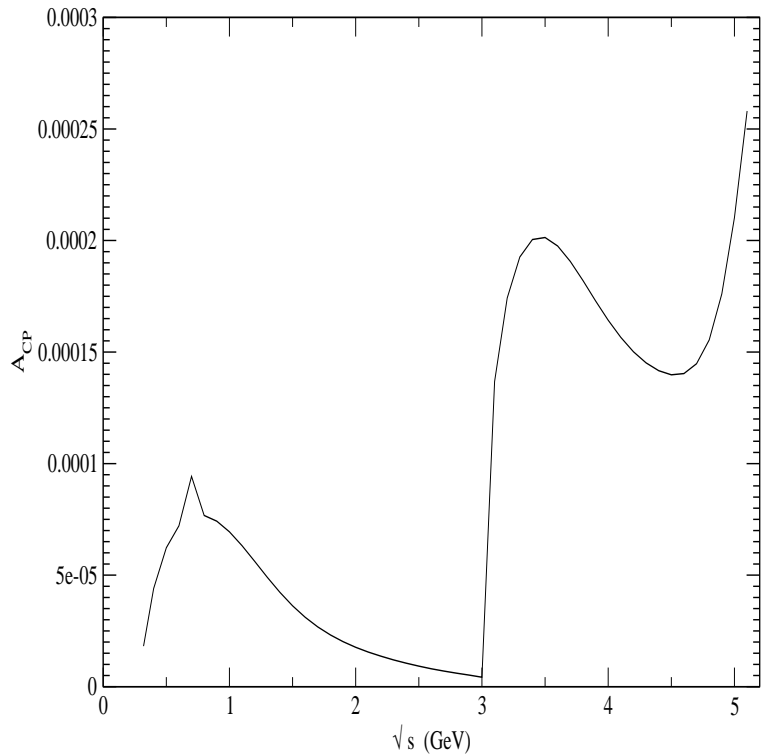}}
\caption{\em The CP asymmetry for $B^-\rightarrow \pi^- \gamma\gamma$}
\label{fig:fig2}
\end{figure}

Figure 2 shows the CP asymmetry, ${\mathcal{A}}_{CP}$ as a function of
$\sqrt{s_{\gamma\gamma}}$. Also, we present the integrated CP asymmetry value
\bea
{\mathcal{A}}_{CP}\vert_{int} &=&
\frac{\int d\sqrt{s_{\gamma\gamma}}\frac{d\Gamma}{d\sqrt{s_{\gamma\gamma}}} -
\int d\sqrt{s_{\gamma\gamma}}\frac{d\bar{\Gamma}}{d\sqrt{s_{\gamma\gamma}}}}
{\int d\sqrt{s_{\gamma\gamma}}\frac{d\Gamma}{d\sqrt{s_{\gamma\gamma}}} +
\int d\sqrt{s_{\gamma\gamma}}\frac{d\bar{\Gamma}}{d\sqrt{s_{\gamma\gamma}}}}
\\ \nonumber
&=& 1.739 \times 10^{-4}
\eea
Experimental observation of this decay can thus lead to a better understanding
of the deeper structure of the effective Hamiltonian and is also expected
to shed light on the possible determination of the relative signs
between various interference terms. The current fluxes at the B-factories 
are of course too small for any possible measuremnt of these numbers, but 
hopefully in the future such measrements will be possible.
\end{section}

\section*{Acknowledgements}
NM would like to thank the University Grants Commission,
India, for financial support during the initial period of the work. 
SRC would like to acknowledge support from
DST, Government of India, under the SERC scheme. This research was also 
supported by the Australian Research Council.
\begin{section}*{Appendix}
We give the input parameters used in the numerical calculations.\\
\begin{table}[ht]
\begin{center}
\begin{tabular}{|l|l|l|l|l|l|l|l|}\hline
$C_1$&$C_2$&$C_3$&$C_4$&$C_5$&$C_6$&$C_7$&$C_8$\\ \hline
-0.222&1.09&0.010&-0.023&0.007&-0.028&-0.301&-0.144 \\ \hline
\end{tabular}
\caption{The approximate values of $C_i's$ at $\mu=m_b$}
\end{center}
\end{table}
\[
m_b=4.8~GeV \hskip 1cm m_c=1.5~GeV \hskip 1cm m_t=175~GeV
\]
\[
m_s=0.15~GeV \hskip 1cm m_u =m_d =0
\]
\[
m_{B^0}=5.2792~GeV \hskip 1cm \Gamma^{B^0}_{\rm{total}}=4.22\times 10^{-13}~GeV
\]
\[
m_{B^+}=5.2789~GeV \hskip 1cm \Gamma^{B^+}_{\rm{total}}=4.21\times 10^{-13}~GeV
\]
\[
m_{\eta_c}=3~GeV \hskip 0.5cm B_{\eta_c}=2.74\times 10^{-3}~GeV^{-1}
\hskip 0.5cm \Gamma_{\rm{total}}^{\eta_c}=1.3\times 10^{-2}~GeV \hskip 0.5cm
f_{\eta_c}=0.35~GeV
\]
\[
m_{\eta}=0.547~GeV \hskip 0.5cm m_{\eta^{\prime}}=0.95778~GeV
\]
\[
B_{\eta}=13.254\times 10^{-3}~GeV^{-1}
\hskip 0.5cm \Gamma_{\rm{total}}^{\eta}=1.18\times 10^{-6}~GeV \hskip 0.5cm
f_{\eta}=-2.4\times 10^{-3}~GeV
\]
We follow the Wolfenstein parametrization of the CKM matrix with
\[
A=0.8 \hskip 1cm \lambda=0.22 \hskip 1cm \eta=0.34 \hskip 1cm \rho=-0.07
\]
\[
V_{tb}\sim 1 \hskip 1cm V_{ts}=-A\lambda^2
\]
\[
V_{cb}=A\lambda^2 \hskip 1cm V_{cs}=1-\frac{\lambda^2}{2}
\]
\end{section}

\end{document}